# Applicability of Telemedicine in Bangladesh : Current Status and Future Prospects

Ahasanun Nessa, M. A. Ameen, Sana Ullah,Kyung Sup Kwak
*Inha University, Incheon, South korea*
anessa_2006@yahoo.com

## Abstract

*Telemedicine refers to the use of information and communication technology to provide and support health care mainly for the purpose of providing consultation. It is also a way to provide medical procedures or examinations to remote locations. It has the potential to improve both the quality and the access to health care services delivery while lowering costs even in the scarcity of resources. Understanding the potentiality of telemedicine, many developing countries are implementing telemedicine to provide health care facility to remote area where health care facilities are deficient. Bangladesh is not an exception to this either. In this paper we mention the reasons why Bangladesh has to move for telemedicine. We also present the past and on-going telemedicine activities and projects in Bangladesh. Analyzing these projects we have found out some factors which should be assessed carefully for successful implementation of telemedicine application. Finally we propose a prototype telemedicine network for Bangladesh that can improve health facilities through telemedicine by making a connection between rural health facility providers and special hospitals.*

## 1. Introduction

Recent advances of Information and communication technologies(ICTs) help society to quickly access enable services for economic and social development. Telehealth /Telemedicine/E-health is the blessing of ICT and is possibly the most prominent of e-business service that can have a major visible effect on the development of society. The concept of telemedicine is not new. Telemedicine has been used since 1959, when a two way video conferencing link was established using microwaves between university o f Nebraska Medical School and state mental hospital [1]. Then the National Aeronautics and Space Administration (NASA) played an important part in the early development of telemedicine. NASA's effort in telemedicine began in the early 1960s when human began flying in space. In the 1970s and 1980s, telemedicine experiments focused on the transmission of medical images using television. In the 1990's the rapid growth of computer and information technology gave a rebirth of telemedicine. Innovation of new technologies enables telemedicine to grow up into more complex and feature-rich technology [2].

In general sense where medical care relies on the face to face communication between patients and doctors, in telemedicine concept physicians treat a patient who is some distance away. The primary purpose of telemedicine is to reach health care service to patient who is some way isolated from specialized care. Telemedicine provides services on 24 hours a day and seven days a week basis. The patient may be living in remote place like rural area or in a ship in deep ocean and even in space craft. In countries where access to medical services is restricted by distance and poor transportation and where health care services are inadequate, telemedicine offers a great opportunity and possibilities to distribute medical services by utilizing ICTs.

Many developing country can not provide minimal health service to population, due to insufficient number of doctors and health care professionals and medical services. Sometimes it is seen, there are clinics and hospital but they are often ill-equipped and especially, outside urban area beyond the reach of normal communications. The inadequate infrastructure makes it more difficult to provide health care in rural and remote areas in right time. If travel costs of a patient to visit a medical specialist are higher than the cost of providing a tele-medical consultation, then telemedicine might be an economically affordable solution.

Bangladesh is one of the densely populated developing countries where most people are living in villages. There is a huge disparity in health care distribution in rural and urban areas. It is also suffering by lack of medical expertise and health care facilities.

In this scenario utilizing its limited resources telemedicine may be a easiest and cheapest way to disseminate health facility to the rural Bangladesh. Telemedicine activities emerged in Bangladesh in mid 1999. The productivity and usability of telemedicine data depends on the availability of high bandwidth. Last few years Information and communication infrastructure of Bangladesh have been experienced huge booming in development. Bangladesh government has given immense importance to ICT for development for economic growth and poverty reduction. In April 2007 Bangladesh got connected to the submarine cable network as a member of the SEA-ME-WE-4 Consortium. Several private and public telecommunication operators have established their network all over the country. As they are expanding their operation to the most rural areas, they are also dwelling to improve the network performance and inclined to introduce latest technologies to the people. Internet facility is almost available in every district of Bangladesh. If government and private organization take proper initiative then it is not so far when a patient in remote places will consult the doctors over internet.

We tried to present the overall telemedicine application and status of Bangladesh in this paper. The organization of this paper is outlined as follows. In Section 2, we present current health care facilities of Bangladesh and recommended telemedicine may be a effective way to distribute health care among urban and rural people by using its limited resources. In Section 3, we present telemedicine activities in Bangladesh till now. Section 4 has been used for discussion by analyzing the past and ongoing telemedicine projects in Bangladesh. We present some factors found from our analysis which are tempting and should be considered for dissemination of telemedicine. In section 5, we propose a prototype telemedicine network for in Bangladesh perspective. Finally we have concluded our paper in section 6.

## 2. Telemedicine in Bangladesh : Why?

Bangladesh is one of the most densely populated countries of the world. About 140 million people living within 144,000 sq. km of land (1045 person/km² ). There are only 663 Government hospitals in district head-quarters and thana (sub-town) areas. Total number of beds available in both public and private hospitals and clinics is 51,648.So the ratio of one hospital bed to citizen of Bangladesh is around 1:2571.

Table-1 gives an overview of existing health care facility of Bangladesh. From this table it is easily visible the inadequate healthcare infrastructure of this highly populated country. Around 80% of the total population of this country lives in rural areas. And rural health centers are often ill-equipped for proper medical treatment. Moreover most of the doctors are city based. After being selected as a cadre of Bangladesh Civil Service (BCS) usually get employment in remote health

Table 1
Health-Care Facilities in Bangladesh

| Facilities | Quantity | Year |
|---|---|---|
| Number of hospital beds | 51,648 | 2005 |
| Population per hospital bed | 2571 | 2005 |
| Hospital beds per 10,000 population | 3.43 | 2005 |
| Number of health centres | 1385 | 2004 |
| Number of physicians | 42,881 | 2005 |
| Population per physician | 3169 | 2005 |
| Physicians per 10,000 population | 3 | 2005 |
| Population per nurses | 6442 | 2005 |

centre of Bangladesh. Due to poor infrastructure of rural health centre and poor infrastructure of villages most of them leave the rural areas within 1-2 years and shift to city area. They feel that they become professionally isolated and outdated if stationed in remote areas. As a result health staffs in rural areas are usually young, have little work experience and show high job rotation. In many cases rural Health Centers are headed by infirmary technicians who are barely trained. In this situation rural people rarely get any specialist doctor's advice when they go to health centers in thana or Upazila level. To get better consultancy rural people spend most of their money on travel to visit doctor in urban areas instead of meeting other treatment expanses. Sometimes, it is not possible to transfer a patient to the suburb or to the city on time due to his/her critical health condition and poor communication facilities in those areas.

Under this circumstance to provide health care in rural area there is two option. One is, building hospital in rural areas and also improves the infrastructure so that doctors and others staffs feel convenient to stayat that places. Another one, is to take any initiative so that it is easy to access quality of health care to rural areas. To implement first one needs huge investment and time. So remaining is second one. To implement second one telemedicine is the best away to provide better health care by using maximum utilization of limited resources.

## 3. History of Telemedicine in Bangladesh

Telemedicine in Bangladesh emerged in before 1999.Many Bangladeshi physicians and surgeons were practicing informal tele-consultation with their colleagues in different countries. The early initiatives were sporadic and unorganized and most of them were based in store and forward technologies such as telegram and e-mail basis. A more formal approach was taken only since 1999.

**Time Line of Telemedicine in Bangladesh**

**1999**: In Bangladesh first telemedicine link was established by a charitable trust named Swifne Charitable. It established the link between the Centre for the Rehabilitation of the Paralysed (CRP) in Dhaka (the capital of Bangladesh) and Royal Navy Hospital, Haslar, UK .It was email based. It used a digital camera to capture still images which was then transmitted by email. It was very successful project. An evaluation of the 27 referrals made during the first year of operation showed that tele-consultancy had been useful and cost-effective. Based on the success of the Bangladesh project, the Swinfen Charitable Trust supplied: digital cameras and tripods to more hospitals in other developing countries.

**July 1999:** Telemedicine Reference Center Ltd. (TRCL) Dhaka, Bangladesh a private company launched his journey with an ambitious project to set up telemedicine project to help doctor identify diseases early on. Dr Sikder M. Zakir, President & CEO of TRCL said "If we get the diseases in early stage then it is possible to save 20 times more money that are being spend". At beginning TRCL linked 200 specialists who will offer their expert opinion to rural doctor. Although it is a privately funded project, Ministry of Health and Family Welfare of the Government of People's Republic of Bangladesh is playing a significant role in TRCL's rural telehealth care program implementation. TRCL started feasibility study and infrastructure development to establish national and international telemedicine services.

**Mid - 2000:** Grameen Communications took rural tele-health initiatives using wireless technology.

**2001**: TRCL demonstrated telemedicine system in the US Trade Show 2001 in Dhaka using Icare software and normal Internet connection & started test-run of the system between US and Bangladeshi physicians.

**April 2001:** The Bangladesh Telemedicine Association (BTA) is formed. But the lack of government sector the technology remains out of reach to majority of poor people of the country

**2003:** Sustainable Development Network Program (SDNP) Bangladesh began in January 2003. The e-HL project built two network segments (8Km; and 6Km) using point-to-point radio with bi-directional bandwidth of 2 Mbps. Currently, SDNP has four regional nodes in different parts of Bangladesh (Cox's Bazar, Dinajpur and Satkhira), including Mymensingh. These nodes are connected to satellite through VSAT 22(SCPC/MCPC) technology. Under this project every Friday they arrange consultancy and diagnostic support to the physician at remote end, though medical experts at the SDNP head office [5].

**2003:** Bangladesh University of Engineering & Technology (BUET) and Comfort Nursing Home had started a Telemedicine project with the financial collaboration from European Union (EU) via Email. Recently the project is not functional.

**May 2004:** Bangladesh. DNS diagnoses Centre, Gulshan-1 and Comfort Diagnoses &Nursing Home's started a Telemedicine centre. The project was discontinued because of lack of financial viability, patient disinterest and poor market promotion

**2005:** In August 16, 2005 Grameen Telecom (GTC) in cooperation with the Diabetic Association of Bangladesh (DAB) launched telemedicine services, giving patients at Faridpur (one district of Bangladesh) General Hospital access to specialist doctors of their choice in Dhaka. DAB's BIRDEM Hospital Dhaka, was connected via a video conferencing link to DAB's Faridpur General Hospital. Consultations now take place over video conference where patient and doctor see each other on television screen. The cost per consultation for a new patient was 600 BDTaka (10 US$), with repeat patients getting some discount [6].

**2006 November:** Telemedicine Reference centre Ltd. And Grameen phone has started a unique telemedicine Service "HealthLine Dial 789" A GSM infrastructure based call centre for Grameen phone 10 million subscribers. Providing different types of medical information facility, emergency service (SMS based LAB report, ambulance) and real time medical consultation over mobile phone [7].

**2007 September:** The project 'ICT in rural Bangladesh' is working for the development of health care facilities in rural Bangladesh through ICT. The project is funded by SPIDER (The Swidish Program for ICT in Developing Regions) for the period September 2007–December 2009 and is collaboration with Grameen Communications, Bangladesh; Grameen Phone, Bangladesh; Bangladesh Sheikh Mujib Medical University, Bangladesh; and International Institute of Information Technology (IIIT), India. The overall goal of the project is to improve the performance of rural health workers and increase the access to healthcare services for rural poor in the district of Magura in Bangladesh. More information about the project.[8].

## 4. Discussion

The first project of telemedicine, between CRP in Dhaka and Royal Navy Hospital, Haslar, was successful and it encouraged other developing country to take this type of initiative. At beginning the outcome of DAB Pilot Project was satisfactory. In the first three months, there were 52 new patients and 6 returning patients. The number of patient rate was not satisfactory later. The patient rate was 1.3 to 1.5 till 2006. The project is still running but now days no patient is seen for service. Later some technical problem had been arisen that decreased the patient rate. The performance of the camera was not satisfactory and picture would damage in rough weather due to the last end Radio link connection. We think another reason of failure of this project is implementing the project in a city which is very near to Dhaka. Faridpur is very near from Dhaka. It took only 1.5-2 hours to go Dhaka from Faridpur by bas. So the patient may have found it more convenient to come to Dhaka and physically meet doctor of their choice. SDNP telehealth project is running now. They only provide tele-consultation with remote doctor on every Friday that is not enough to support any emergency situation.

Tele-psychiatry is another application of Telemedicine. It is the most successful and low cost telemedicine applications, because it needs only a good videoconferencing between two end points. According to National Mental Health Survey in 2003-2005 about 16.05% of the adult population of Bangladesh is suffering from mental disorders [9]. There is a high prevalence of psychiatric disorders in rural Bangladesh. Money spends for mental services are less than 0.5% of the total national health expenditure. There are 50 outpatient mental health facilities, 31 community-based psychiatric inpatient units, 111 community residential facilities and one 500 bedded mental hospital in the country. In past telemedicine projects tele-psychiatry did not get attention. It can be considered as way for providing mental health care.

Although telemedicine is not a panacea for all problems related of health care, it can help to minimize the problems that are related with time in distant. Telemedicine activities are still in primary level in Bangladesh. There is no support to meet the emergency medical need such as a stroke, severe injury for people in remote areas. Many developing countries have implemented telemedicine networks which have made a connection between remote hospitals and special hospitals. We can take India, China as for example [10]. China has three major telemedicine networks. The IMNC network is primarily based on telephone line and Internet. To possible data transmission over low bandwidth they are using powerful image compression algorithm that can reduce the file size dramatically.[11].

The successful expansion of Telemedicine fully depends on the improvement of Information and telecommunication structure. In past telemedicine implementation was not easy due to poor telecommunication infrastructure. Now the situation is in favor of telemedicine. Telecommunication sector in Bangladesh has been experiencing a huge boom in the last few years. Several private and public telecommunication operators are expanding their operation to the most rural areas; they are also dwelling to improve the network performance and inclined to introduce latest technologies to the people. Government always patronizes past telemedicine projects. Government has lots of responsibility in this aspect. Government should encourage private organization to invest in telemedicine implementation.

Since telemedicine practicing is increasing day by day so it is very much needed to provide structured laws and regulations about physician provided service, patients' issues, licensing of physician and telemedicine providers. There should be clear rules about reimbursement issues. Since in telemedicine system local doctor or paramedic treats a patient and they consult with specialist who is far away, who will be responsible in patient issues? Government also should create policy in favor of rural peoples so that they can enjoy low cost telemedicine services.

The success of any system depends on user acceptance. Though there is no difference between conventional consultation and a tele-consultancy, patient and doctors who are accustomed to personal visit may be reluctant to alter the traditional methods of health care. May be IT savvy medical staffs and physician will face difficulty in using new technology. To overcome this problem, sufficient training programs should be offered to enable physician to adapt this new technology. Media can play a significant role to make telemedicine popular to user. They should broadcast the successful case history considering the efficacy and cost effectiveness of telemedicine.

There are many factors influencing the diffusion of telemedicine as well as other new technology. The following factors should be assessed before considering telemedicine implementation in a specific health care.

**Medical Need:** First we have to consider what kind of medical facilities can be provided by this specific health centre. Establishing a high-end telemedicine cell in thana (sub-town) level would not be significant because of the structure of health care.

**Organization and Structure of Health care:** According to structure of health care, telemedicine support can be different.

**Location of the Health Centre:** If the location of the health centre is in same city of the super or large hospital then people would prefer face to face consultancy rather than tele-consultancy. We faced this problem in DAB project where the health centre is in Faridpur and special hospital is in Dhaka, the capital of Bangladesh. The distance between the two locations is one and half hour by bus.

## 5. Telemedicine Network Prototype

Telemedicine can be divided into two basic modes of operation: real time and store and forward[12]. In real time telemedicine treatment, the patient is accompanied by an attending physician who consults with specialist who is far way from them. In store and forward mode, all relevant information is transmitted electrically to the specialist. For this mode, the response does not have to be immediate. An ideal telemedicine is a combination of these two modes. High bandwidth is very essential for real time telemedicine.

In Bangladesh maximum large hospital and special hospitals are in Dhaka. To make medical care to remote people one way is to establish same kind of medical units equipped with latest medical equipment in rural areas and also provide specialist medical staffs. It requires huge investment of time, finances and effort. Another possible solution is connecting already established health centers with advanced medical establishments of the cities. In current Bangladesh Telecommunication infrastructure real time telemedicine is possible up to district level among special health care providers and district hospitals. There is optical fiber link in most of the districts. Internet facility is also available in maximum sub-town (Thana). But bandwidth is not high. Store and forward basis telemedicine support can be expanded up to thana level by using telephone line.

Considering the communication background of Bangladesh we are proposing a telemedicine network prototype. The objective of our proposed model is to build up a connection between remote health centre and large or special hospitals. It will help remote doctor to diagnosis diseases at the early stage. It will minimize unnecessary patient transfer, from a remote site. In our proposed model, we are categorizing all kinds of health care providers including hospitals, special hospitals and Health centers into three categories. Cat1: large hospital and special hospital. Cat2: district hospitals. Cat 3: thana health complex.

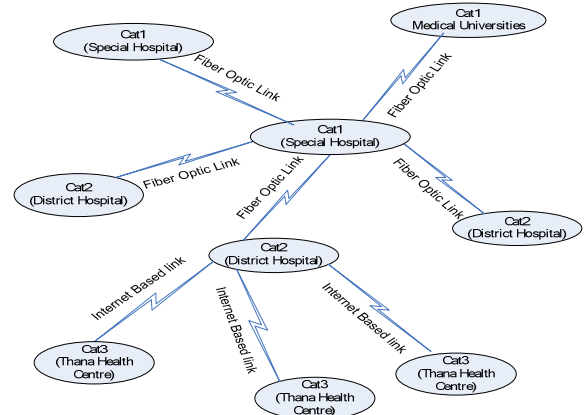

Figure 1. Proposed Telemedicine Network

**Equipment and peripheral**

Both Cat 1 and Cat 2 are equipped with latest telemedicine equipment such as BP monitor, speedometer, ultrasound, tele-ECG, Digital camera, Document camera, glucometer, stethoscope etc. And also need PC and standalone video-conferencing tools. Wireless telemedicine equipment is most appreciable. In Cat1, there is a available team of specialists to provide guidelines, opinion and diagnosis to Cat2 level doctors. Cat 1 is connected with other Cat1s and medical universities.

Cat3 is arranged by medium level of telemedicine equipment so that they can collect patient records and send to special hospital for consultation They refer the difficult cases to Cat2 where the patient gets treatment. Cat2 discuss with Cat1 if any help they needed. They can further refer to Cat1 if necessary.

**Network Backbone:**

In our prototype model Cat1 is connected with Cat2 by optical fiber link. For connection between Cat3 and Cat1 we can use digital data network. Where the duty physician will collect patient record and other reports. After compressing data they send the record to Cat2 hospital if they need any help. With a 28.8 Kbps

Table 2

Our proposed model

| Environment and Setting | Telemedicne mode | Communication backbone | Instrument | Outcomes and results |
|---|---|---|---|---|
| Connetion between Special hospital and district Hospital Connection among special hospitals. | Real time | Fiber optic link | Equipped with latest medical and telemedicine technology, and specialist physician | Provide quick and efficient medical consultation and other services and support |
| Connection between district hospital and Thana health complex | Store and forward | Internet using digital data network | Paramedic or newly employed doctor and medium | Early diagnosis of disease. |

dial-up connection, transmission of a standard X-ray takes 30 minutes. So better to implement store and forward telemedicine technique between Cat3 and Cat2. Table 2 shows our proposed model at a glance.

## 7. Conclusion:

Telemedicine promises for providing significant improvement and cost effective access to quality of health care to under-served communities. In this paper we focus on the necessity of implementing telemedicine application in Bangladesh. We also present the past and ongoing telemedicine project and future prospect of telemedicine with respect to Bangladesh. We strongly feel that the Government should actively patronize private organization for investing in telemedicine sector. Telemedicine will bring benefits not only by providing health care services to remote people but also generating new source of employment.

## Acknowledgment

This research was supported by the MKE(Ministry of Knowledge Economy), Korea ,under the ITRC( Information Technology Research Center ) support program supervised by the IITA( Institute of Information Technology Assessment )" (IITA-2008-C1090-0801-0019)